\documentclass[sigconf,screen]{acmart}
\usepackage{multirow}

\AtBeginDocument{%
  \providecommand\BibTeX{{%
    \normalfont B\kern-0.5em{\scshape i\kern-0.25em b}\kern-0.8em\TeX}}}

\setcopyright{acmcopyright}
\copyrightyear{2023}
\acmYear{2023}
\setcopyright{acmlicensed}\acmConference[RecSys '23]{Seventeenth ACM Conference on Recommender Systems}{September 18--22, 2023}{Singapore, Singapore}
\acmBooktitle{Seventeenth ACM Conference on Recommender Systems (RecSys '23), September 18--22, 2023, Singapore, Singapore}
\acmPrice{15.00}
\acmDOI{10.1145/3604915.3608787}
\acmISBN{979-8-4007-0241-9/23/09}

\begin{document}

\title{Knowledge-based Multiple Adaptive Spaces Fusion for Recommendation}



\author{Meng Yuan}
\affiliation{%
  \institution{Institute of Artificial Intelligence, Beihang University}
  \city{Beijing}
  \country{China}}
\email{yuanmeng97@buaa.edu.cn}

\author{Fuzhen Zhuang}
\authornote{Corresponding author.}
\authornote{Fuzhen Zhuang is also at Zhongguancun Laboratory, Beijing, China}
\affiliation{%
  \institution{Institute of Artificial Intelligence, Beihang University}
  \city{Beijing}
  \country{China}
}
\email{zhuangfuzhen@buaa.edu.cn}

\author{Zhao Zhang}
\affiliation{%
 \institution{Institute of Computing Technology, Chinese Academy of Sciences}
 \city{Beijing}
 \country{China}}
\email{zhangzhao2021@ict.ac.cn}

\author{Deqing Wang}
\affiliation{%
  \institution{School of Computer Science and Engineering, Beihang University}
  \city{Beijing}
  \country{China}}
\email{dqwang@buaa.edu.cn}

\author{Jin Dong}
\affiliation{%
  \institution{Beijing Academy of Blockchain and Edge Computing}
  \city{Beijing}
  \country{China}}
\email{dongjin@baec.org.cn}


\begin{abstract}
  Since Knowledge Graphs (KGs) contain rich semantic information, recently there has been an influx of KG-enhanced recommendation methods. Most of existing methods are entirely designed based on euclidean space without considering curvature. However, recent studies have revealed that a tremendous graph-structured data exhibits highly non-euclidean properties. Motivated by these observations, in this work, we propose a knowledge-based multiple adaptive spaces fusion method for recommendation, namely MCKG. Unlike existing methods that solely adopt a specific manifold, we introduce the unified space that is compatible with hyperbolic, euclidean and spherical spaces. Furthermore, we fuse the multiple unified spaces in an attention manner to obtain the high-quality embeddings for better knowledge propagation. In addition, we propose a geometry-aware optimization strategy which enables the pull and push processes benefited from both hyperbolic and spherical spaces. Specifically, in hyperbolic space, we set smaller margins in the area near to the origin, which is conducive to distinguishing between highly similar positive items and negative ones. At the same time, we set larger margins in the area far from the origin to ensure the model has sufficient error tolerance. The similar manner also applies to spherical spaces. Extensive experiments on three real-world datasets demonstrate that the MCKG has a significant improvement over state-of-the-art recommendation methods. Further ablation experiments verify the importance of multi-space fusion and geometry-aware optimization strategy, justifying the rationality and effectiveness of MCKG.
\end{abstract}

\ccsdesc[500]{Information systems~Recommender systems}

\keywords{Knowledge Graph, Recommender Systems, Multiple Space Fusion, Geometry-aware Optimization Strategy}


\maketitle

\section{Introduction}
Recommender systems (RSs) have shown great potential in solving the information explosion problem and enhancing the user experience in various online applications~\cite{sun2020multi}. In a variety of scenarios, knowledge graphs (KGs) can be utilized to offer fundamental background knowledge as well as rich structural information. To fully utilize KG, existing knowledge-enhanced recommendation methods ~\cite{qu2019end,huang2018improving,wang2018ripplenet,wang2021learning, chen2020revisiting} strive to build more effective neural networks to integrate the semantics of KG. 

\begin{figure}
\centering
  \includegraphics[scale=.9, trim={12mm 0mm 0mm -6mm}]{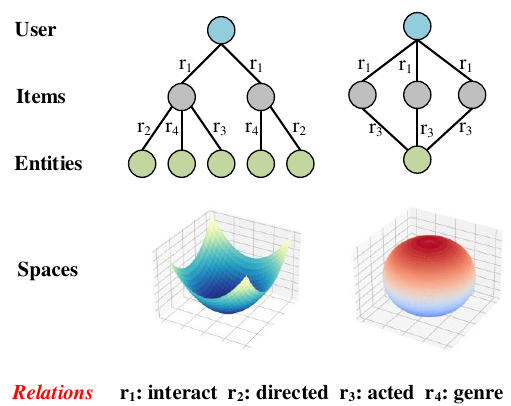}
    \caption{Tree and cyclic structures correspond to their most suitable modeling spaces: the hyperbolic and spherical spaces.}
    \label{Fig1}
\end{figure}

Although these techniques can significantly enhance the embeddings of users and items, they are entirely designed based on Euclidean space without considering curvature. Recent studies ~\cite{bachmann2020constant, peng2021hyperbolic,zhang2022geometric} have revealed that a tremendous data exhibits the highly non-Euclidean properties, especially the data that presents tree-likeness or cyclic structures. For instance, the KG data are diverse and include a variety of structures. As shown in Figure~\ref{Fig1}, there are interactions between the user and items, which can be further connected to more entities. Owing to such high-order relationship~\cite{wang2019kgat}, knowledge-aware methods have the ability to provide accurate recommendations. Essentially, such interactions can be seen as tree structures or cyclic structures. In such situations, Euclidean space methods suffer from severe distortion when depicting these structures~\cite{chami2019hyperbolic, sala2018representation}. Furthermore, recent studies~\cite{liu2019hyperbolic, gu2019learning} have demonstrated that geometric spaces with constant non-zero curvature can enhance representation performance when the underlying graph structures of the data follow particular patterns. Specifically, the hyperbolic space is more suitable for modeling tree-structured data, while the spherical space is superior for representing cyclic patterns.

Since complex KG usually have varied structures coupled together, choosing the ideal space is especially tricky. Constant curvature space methods~\cite{chen2022modeling, tai2021knowledge, bachmann2020constant, grattarola2018learning} just project entities to a single space, as a result, they fail to completely express the sophisticated structures of KG. Recently, some mixed-curvature space works~\cite{gu2018learning, han2020dyernie, wang2021mixed} merge distinct spaces, aiming to characterize various structures. Unfortunately, how to find the most effective combination of these manifolds is also a challenging task. The simplest method~\cite{zhang} is to list all possible combinations and choose the one that best utilizes these geometry. Obviously, this method would be efficient but impractical due to the time-consuming and low-scalability~\cite{skopek2019mixed}. Another solution~\cite{chen2022modeling} is to treat the curvature $\kappa$ as a trainable parameter, aiming to learn the optimal curvature from the training data. However, there are still the following issues: 1) the constraints of model parameters depend on the changes of curvature. Once the curvature $\kappa$ varies (e.g., from negative to positive), parameters may no longer satisfy the constraints. As a result, the internal structure of manifold will be destroyed, thus unable to obtain high-quality embeddings. 2) Since the manifolds belong to different spaces, i.e., $\mathcal{M} \in \mathbb{H}, \mathbb{E}, \mathbb{S}$, directly operating on various manifolds without considering the heterogeneity is problematic. For instance, the global distance computation just sums up the distances without distinguishing the importance of various manifolds. 

Motivated by the above observations, in this paper, we propose a knowledge-based multiple adaptive spaces fusion method for recommendation (MCKG). Unlike existing methods~\cite{chen2022modeling, tai2021knowledge,sun2021hgcf, yang2022hicf} that solely consider a specific manifold, we introduce a unified space $\mathbb{U}$ that is compatible with the hyperbolic, euclidean and spherical spaces. Note that the unified space can interpolate smoothly between all geometries of constant curvature, thus the internal structure of the manifold is preserved when training curvature. To break the limitation of the single space's expression ability, we further integrate multiple unified spaces to more accurately capture the structural information on a global level. On the other hand, we propose a geometry-aware optimization strategy that enables the pull and push processes benefited from both hyperbolic and spherical spaces. Specifically, in hyperbolic space, we set smaller margins in the area near to the origin, which is conducive to distinguishing between highly similar positive items and negative ones. At the same time, we set larger margins in the area far from the origin to ensure that the model has sufficient error tolerance. Conversely, in spherical space, the geometry-aware strategy assigns larger margins near to the origin and smaller margins far from the origin. Empirically, we conduct comprehensive experiments on the three benchmark datasets, the results show that MCKG outperforms the state-of-the-art recommendation methods. In summary, this work makes the following contributions:
\begin{itemize}

\item  To the best of our knowledge, this is the first work to apply multiple adaptive spaces fusion for knowledge-enhanced recommender systems.

\item  We present a knowledge-based multiple adaptive spaces fusion method for recommendation. To obtain the high-quality embeddings for recommendation, we introduce the unified space to describe the complex structures of KG, then further fuse multiple subspaces in an attention manner. Finally, our proposed geometry-aware optimization strategy is the first work that considers the properties of both hyperbolic and spherical spaces.

\item  We conduct extensive experiments on the three benchmark datasets, results show that MCKG outperforms the state-of-the-art recommendation methods. 
\end{itemize}

\section{PRELIMINARIES AND PROBLEM FORMULATION}

\subsection{Mathematical Concepts}
In this subsection, we summarize preliminary notations for readers to better understand this paper.

\textbf{Manifold}: A manifold $\mathcal{M}$ of dimension $n$ is a topological where each point's neighborhood can be approximated by euclidean space $\mathbb{R}^{n}$. For example, the earth can be modeled by the spherical space, and its local place can be estimated by $\mathbb{R}^{2}$. The notion of manifold is a generalization of the notion of surface. 

\textbf{Tangent space}: For each point $x\in\mathcal{M}$, the tangent space $\mathcal{T}_{x}\mathcal{M}$ of $\mathcal{M}$ at $x$ is defined as a $n$-dimensional space estimating $\mathcal{M}$ around x at a first order. 

\textbf{Geodesics distance}:  Geodesics distance is the the generalization of a straight line in the Euclidean space, which indicates the shortest path between pairs of points.

\textbf{Exponential map}: The exponential map carries a vector $v\in\mathcal{T}_{x}\mathcal{M}$ of a point $x\in\mathcal{M}$ to the manifold $\mathcal{M}$, i.e., Exp$_{x}: \mathcal{T}_{x}\mathcal{M}\rightarrow \mathcal{M}$ by traveling a fixed distance along the geodesic defined as $\gamma(0)=x$ with direction $\gamma'(0)=v$. Each manifold has its own manner to design the exponential maps.

\textbf{Logarithmic map}: The logarithmic map is the reverse operation of the exponential map, which projects a point $z \in \mathcal{M}$ on the manifold back to the tangent space of another point $x \in \mathcal{M}$, i.e., Log$_x:\mathcal{M} \rightarrow \mathcal{T}_{x}\mathcal{M}$. Similar to Exp$_{x}$, different manifolds correspond to their distinct logarithmic map.

\subsection{Constant Curvature Spaces}
With the constant sectional curvature $\kappa$, the three types of manifolds are defined as follows:
\begin{small}
\begin{equation}
\begin{split}
\hspace{-6mm}
\qquad {\mathcal{M}_{\kappa}^{n}}&= \left \{
\begin{array}{ll}
{{\mathbb{H}_{\kappa}^{n}:}\{x \in \mathbb{R}^{n+1}:\langle x, x\rangle _{\kappa} = 1/\kappa, x_{0} > 0 \}} & for\ \kappa < 0\\[2mm]
{\mathbb{E}_{\kappa}^{n}:\mathbb{R}^{d}} & for\ \kappa = 0\\[2mm]
{{\mathbb{S}_{\kappa}^{n}:}\{x \in \mathbb{R}^{n+1}:\langle x, x\rangle _{\kappa} = 1/\kappa \}}& for\ \kappa > 0
\end{array}\right.
\end{split}
\label{eq:Eq_1}    
\end{equation}
\end{small}

where the $\langle \cdot, \cdot \rangle_{\kappa}$ represents the inner product:
\begin{equation}
\begin{split}
\hspace{-6mm}
\qquad {\langle x,y \rangle_{\kappa}}&= \left \{
\begin{array}{ll}
{\sum_{i=0}^{n}x_{i}y_{i}} & for\ \kappa > 0\\[2mm]
{-x_{0}y_{0}+\sum_{i=1}^{n}x_{i}y_{i}}& for\ \kappa < 0
\end{array}\right.
\end{split}
\label{eq:Eq_2}    
\end{equation}
Because hyperbolic or spherical spaces are not vector spaces, the vector operations (e.g., addition, subtraction and scalar multiplication) cannot be achieved. Therefore, we need to convert non-Euclidean spaces into tangent spaces for corresponding vector calculations. Specifically, the tangent space $\mathcal{T}_{x}\mathcal{M}$ at point $x$ on $\mathcal{M}$ is a $d$-dimensional Euclidean space that approximates $\mathcal{M}$ around $x$:
\begin{equation}
\begin{split}
\mathcal{T}_{x}\mathcal{M}_{\kappa}^{m} = \{v\in \mathbb{R}^{n+1} :\langle v, x\rangle_{\kappa}=0\}.
\end{split}
\label{eq:Eq_3}    
\end{equation}
As mentioned in the previous section, the mapping between manifold $\mathcal{M}_{\kappa}^{n} \in \{\mathbb{H}_{\kappa}^{n}, \mathbb{S}_{\kappa}^{n} \}$ and its tangent space $\mathcal{T}_{x}\mathcal{M}_{\kappa}$ can be achieved by the exponential  and logarithmic operations. 

To connect vectors in tangent spaces, the parallel transport~\cite{helgason1978differential} defines a manner of transporting the local geometry along smooth curves that preserves the metric tensors. Specifically, for vector $u \in \mathcal{M}$ and $v \in \mathcal{M}$, the parallel transport $\mathcal{P}\mathcal{T}_{u\rightarrow v}:\mathcal{T}_{u}\mathcal{M}\rightarrow \mathcal{T}_{v}\mathcal{M}$ carries a vector in $\mathcal{T}_{u}\mathcal{M}$ along the geodesic from $u$ to $v$.

\subsection{Mix-Curvature Spaces}
Since the constant curvature space is only suitable for a certain geometric structure, mix-curvature methods~\cite{bachmann2020constant, gu2018learning, zhang, Kochurov2020, skopek2019mixed} expect to match the complicated geometry of data thus provide higher quality representations. 
To capture a wider range of curvatures, mix-curvature methods propose embeddings into product spaces~\cite{gu2018learning} where each component has constant curvature. Formally, assuming a series of $N$ distinct constant curvature spaces $\mathcal{M}^{(1)}, \mathcal{M}^{(2)}, ..., \mathcal{M}^{(N)}$, the mix-curvature manifold are defined as: 
\begin{equation}
\begin{split}
\mathcal{M} = \mathcal{M}^{(1)} \times \mathcal{M}^{(2)} \times ... \times \mathcal{M}^{(N)},
\end{split}
\label{eq:Eq_4}    
\end{equation}
where the  $\times$ denotes Cartesian product.

\subsection{Problem Formulation}
We follow the recommendation setting~\cite{wang2018ripplenet, wang2019kgat, chen2022modeling}.  Let $U=\{u_1,u_2,...\}$ and $V=\{v_1,v_2,...\}$ denote the sets of users and items, respectively. The user-item interaction matrix $Y=\{y_{uv} | u \in U,v\in V\}$ is defined based on the users' implicit feedback, where the value of $y_{uv}=1$ denotes that there is an interaction between user $u$ and item $i$; otherwise $y_{uv} = 0$. We also have a knowledge graph $G$ available, which consists of massive entity-relation-entity triplets $(h,r,t)$, where $h\in \phi$, $r\in \varphi$, and $t\in \phi$ describe the head, relation, and tail of knowledge triples, and $\phi$ and $\varphi$ denote the set of entities and relations. Given the user-item interactions and the item-side KG, the recommendation task is to train a RS model calculating the probability that the user $u$ will click item $v$.

\section{METHOD}
\begin{figure*}
\centering
    \includegraphics[scale=.73, trim={5mm 0mm 0mm 0mm}]{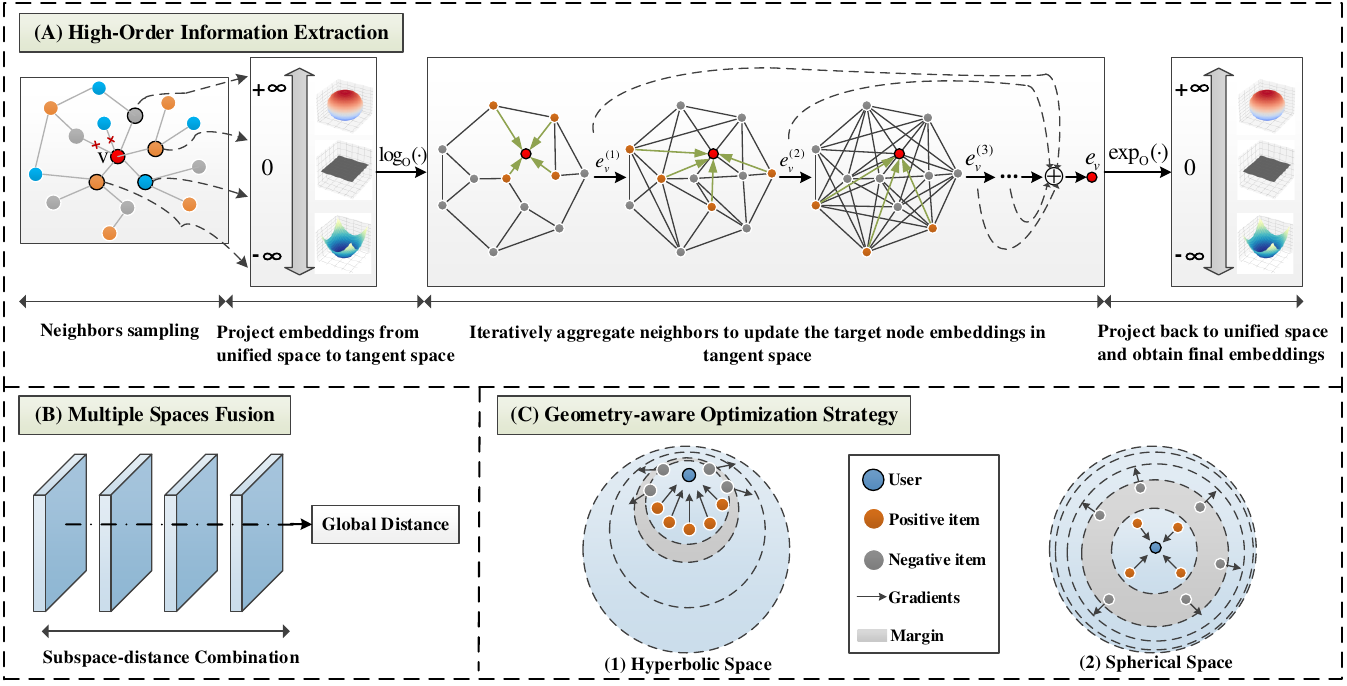}
    \caption{Overview of MCKG model architecture: (A) The overall process of high-order information extraction based on the single adaptive space $\mathbb{U}_{\kappa}^{n}$; (B) Aggregating the subspace embeddings and adopt an attention mechanism to calculate the global distance. (C) The proposed optimization strategy enables the pull and push processes to be geometric-aware in both hyperbolic and spherical spaces.}
    \label{Fig2}
\end{figure*}

In this section we present the MCKG model in Figure~\ref{Fig2}. First, we introduce the $\kappa$-Stereographic manifold and the extraction of high-order information on a single manifold. Then, we fuse the each adaptive subspaces to fully capture the global structural information. Finally, we propose a novel geometry-aware optimization strategy to be compatible with both hyperbolic and spherical spaces.

\begin{table}
\renewcommand{\arraystretch}{1}
\caption{\textbf{Summary of operations in unified space $\mathbb{U}_{\kappa}^{n}$}}
\centering
\begin{tabular}{lc}
\hline
Name & Operation \\
\hline
Addition & $x \oplus_{\kappa} y = \frac{(1-2\kappa \langle x, y\rangle - \kappa \|y\|_{2}^{2})x+(1+\kappa \|y\|_{2}^{2})y}{1-2\kappa \langle x, y\rangle + \kappa^{2}\|x\|_{2}^{2}\|y\|_{2}^{2} }$ \\[3mm]
Multiplication & $x \otimes_{\kappa} y = \operatorname{exp}_{\mathbf{o}}^{\kappa}(x \cdot \operatorname{log}_{\mathbf{o}}^{\kappa}(y))$\\[3mm]
Concatenate & $x \ominus_{\kappa} y = \operatorname{exp}_{\mathbf{o}}^{\kappa}(x || \operatorname{log}_{\mathbf{o}}^{\kappa}(y))$\\[3mm]
Dot Product & $x \odot_{\kappa} y = \operatorname{exp}_{\mathbf{o}}^{\kappa}(x^{T} \operatorname{log}_{\mathbf{o}}^{\kappa}(y))$\\[3mm]
Geodesics Distance & $d_{\kappa}(x,y) = 2\operatorname{tan}_{\kappa}^{-1}(\|-x\oplus_{\kappa} y\|_{2})$\\[3mm]
Exponential Map & $\operatorname{exp}_{x}^{\kappa}(v) = x \oplus_{\kappa}(\operatorname{tan}_{\kappa}(\frac{\lambda_{x}^{\kappa}\|v\|_{2}}{2})\frac{v}{\|v\|_{2}}$\\[3mm]
Logarithmic Map & $\operatorname{log}_{x}^{\kappa}(y) = \frac{2}{\lambda_{x}^{\kappa}}\operatorname{tan}_{\kappa}^{-1}(\|-x\oplus_{\kappa}y\|_{2})\frac{-x\oplus_{\kappa}y}{\|-x\oplus_{\kappa}y\|_{2}}$\\[3mm]
\hline
\end{tabular}
\label{tab1}
\end{table}

\subsection{Unifying All Curvatures}
We adopt $\kappa$-Stereographic model ~\cite{bachmann2020constant} as the base manifold, which is a unification of constant curvature manifolds: ${\mathbb{H}_{\kappa}^{n}}$, ${\mathbb{E}_{\kappa}^{n}}$ and $  {\mathbb{S}_{\kappa}^{n}}$. The model is a Riemannian manifold with constant sectional curvature $\kappa$ and dimension $n$: 

\begin{equation}
\begin{split}
\hspace{-6mm}
\qquad {\mathbb{U}_{\kappa}^{n}}&= \left \{
\begin{array}{ll}
{{x \in \mathbb{R}^{n}: \|x\|_{2}<1/\sqrt{-\kappa}}}, \lambda_{x}^{\kappa} & \ \kappa < 0\\[1mm]
{\mathbb{R}^{n}}, \lambda_{x}^{\kappa} & \ \kappa \geq 0\\[1mm]
\end{array}\right.
\end{split}
\label{eq:Eq_5}    
\end{equation}

where $\lambda_{x}^{\kappa}=\frac{2}{1+\kappa \|x\|_{2}^{2}}$ is conformal metric tensor at point $x$. For the $\kappa$-Stereographic model, the optimization step changes the geometry of space and the constraints of parameters, which also depend on the curvature. Fortunately, when we use curvature $\kappa$ as a trainable parameter, the model can interpolate smoothly between all geometries of constant curvature, and the 
detailed proof process can refer to paper ~\cite{bachmann2020constant}. 

The basic operations of the unified space ${\mathbb{U}_{\kappa}^{n}}$ are summarized in Table \ref{tab1}, but the trigonometric functions in Table~\ref{tab1} are not complete, here we complement its definition.
\begin{equation}
\begin{split}
\qquad {tan_{\kappa}(x)}&= \left \{
\begin{array}{lr}
{\kappa^{-1/2} tan(x\kappa^{1/2})} & \ \kappa > 0\\[2mm]
{x+\frac{1}{3}\kappa x^{3}}& \ \kappa = 0\\[2mm]
|\kappa|^{-1/2}tanh(x|\kappa|^{1/2})& \kappa < 0\\
\end{array}\right.
\end{split}
\label{eq:Eq_20}    
\end{equation}

\begin{equation}
\begin{split}
\qquad {tan_{\kappa}^{-1}(x)}&= \left \{
\begin{array}{lr}
{\kappa^{-1/2} tan^{-1}(x\kappa^{1/2})} & \ \kappa > 0\\[2mm]
{x-\frac{1}{3}\kappa x^{3}}& \ \kappa = 0\\[2mm]
|\kappa|^{-1/2}tanh^{-1}(x|\kappa|^{1/2})& \kappa < 0\\
\end{array}\right.
\end{split}
\label{eq:Eq_3}    
\end{equation}

\subsection{High-Order Information Extraction}
Since the initial embeddings are in Euclidean space, we explicitly encode the Euclidean embeddings onto the unified space before feeding the subsequent layers. Denote the initial user, item and relation embeddings as $x_u$, $x_v$ and $x_r$ respectively, then the exponential maps are applied:
\begin{equation}
\begin{split}
e_{u}=exp_{\mathbf{o}}^{\kappa}(x_u),\quad e_{v}=exp_{\mathbf{o}}^{\kappa}(x_v),\quad
e_{r}=exp_{\mathbf{o}}^{\kappa}(x_r).
\end{split}
\label{eq:Eq_6}    
\end{equation}

We present a relational attention function as follows to describe user's interest:

\begin{equation}
\begin{split}
 \quad {c}_{{u}}^{{r}} = {e_u}\odot_{\kappa} {e_r}.
\end{split}
\label{eq:Eq_7}
\end{equation}

Intuitively, ${c}_{{u}}^{{r}}$ means the importance of relation $r$ to user $u$. The neighboring information can be continuously aggregated based on $c_{u}^{r}$. Specifically, the neighborhood embedding of entity $v$ is defined as 
\begin{equation}
\begin{split}
{e_{s(v)}^{u} = \sum \limits_{a\in s(v)} {\hat {c}}_{u}^{r}\otimes_{\kappa} e_a}, \quad {\hat {c}}_{u}^{r}= \frac{Exp(c_{u}^{r_{v,a}})}{\sum \limits_{a\in s(v)}  Exp(c_{u}^{r_{v,a}})},
\end{split}
\label{eq:Eq_8} 
\end{equation}
where $s(v)$ denotes the selected neighbors of $v$, the ${\hat {c}}_{u}^{r}$ is the normalized user-relation score, and $Exp(\cdot)$ means the basic exponential function.

The key idea of Graph Convolution Network (GCN) is to aggregate feature information from a node's neighbors. Obviously, feeding the entire knowledge graph to model will greatly increase the computational burden. To this end, we assign a fixed size receptive field for each node to control the amount of calculation. After getting the neighborhood representation, we aim to represent target node at the next layer. There are three kinds of aggregators:

\begin{itemize}
\item GCN Aggregator~\cite{niepert2016learning} adds the two representations and performs a nonlinear transformation, and it can be formulated as
\begin{equation}
\begin{split}
{{e_v^{(k+1)}}=exp_{v}^{\kappa}(\sigma log_{v}^{\kappa}(W\otimes_{\kappa}(e_{v}^{(k)}\oplus_{\kappa}e_{s(v)}^{(k)})\oplus_{\kappa}b)}), 
\end{split}
\label{eq:Eq_9}
\end{equation}
where $\sigma$ is LeakyRelu activation function, $W$ and b represent the weight matrix and bias respectively, $\otimes_{\kappa}$ and $\oplus_{\kappa}$ denote the multiplication and addition as mentioned in Table~\ref{tab1}. 

\item GraphSage Aggregator~\cite{hamilton2017inductive} concatenates the two representations, then fed to the activation function:
\begin{equation}
\begin{split}
{{e_v^{(k+1)}}=exp_{v}^{\kappa}(\sigma log_{v}^{\kappa}(W\otimes_{\kappa}(e_{v}^{(k)}\ominus_{\kappa} e_{s(v)}^{(k)})\oplus_{\kappa}b)}),
\end{split}
\label{eq:Eq_10} 
\end{equation}
where the $\ominus$ is the concatenation operation.

\item Neighbor Aggregator~\cite{wang2019kgat} takes the neighborhood
representation of entity $v$ as the output representation, and it can be formulated as
\begin{equation}
\begin{split}
{e_v^{(k+1)}}=exp_{v}^{\kappa}(\sigma log_{v}^{\kappa}(W\otimes_{\kappa}(e_{s(v)}^{(k)})\oplus_{\kappa}b)).
\end{split}
\label{eq:Eq_11}
\end{equation}
\end{itemize}

The embeddings at different layers capture different semantics. E.g., the first layer focuses on items that users has directly interacted with, the second layer emphasizes overlapping items between users, and the higher-layers can capture higher-order information~\cite{wang2019neural}. After $K$ layers aggregation, we further combine the embeddings obtained at each layer to form the final item representations:
\begin{equation}
\begin{split}
 {e_v^{*}}= {e_v^{(0)}}\oplus_{\kappa}{e_v^{(1)}}...\oplus_{\kappa}{e_v^{(k)}},
 {e_u^{*}}= {e_u^{(0)}}.
\end{split}
\label{eq:Eq_12}
\end{equation}

\subsection{Multiple Spaces Fusion}
Since the complicated structures are not evenly distributed over the entire graph, even if a single space can learn the ideal curvature $\kappa$, it is still not capable of fully modeling the entire graph. Now we consider a sequence of manifolds $\mathbb{U}^{(1)}, \mathbb{U}^{(2)},...,\mathbb{U}^{(M)}$, where the item (user) embedding in the $m$-th subspace is denoted as $e_v^{*,m}$. To accurately capture the global structural information, we design a fusion mechanism by aggregating subspace embeddings. First, we take the average of subspace representations as the global fused embeddings ${e_v^{*,g}}$, then we concatenate the average value and the original subspace embedding to obtain the new subspace embedding:

\begin{equation}
\begin{split}
 {e_v^{*,g}}= \frac{1}{M}\sum_{m=1}^{M}{e_v^{*,m}},\\
 e_v^{*,m} = e_v^{*,m} \ominus_{\kappa} {e_v^{*,g}}.\\
\end{split}
\label{eq:Eq_13}
\end{equation}

As a result, the new embeddings take into account the global information, which is not limited to the current space. Finally, we further integrate the series of subspaces in an attention-enhanced manner. In other words, we calculate the global distance in an attention manner. Formally, the global distance between user $u$ and item $v$ is the weighted sum of the distances in each subspace: 
\begin{equation}
\begin{split}
 dist(u,v) = \sum_{m=1}^{M}w({e_u^{*,m}}, {e_v^{*,m}})d_{\kappa}({e_u^{*,m}},{e_v^{*,m}}), 
\end{split}
\label{eq:Eq_15}
\end{equation}
where the $d_{\kappa}(\cdot)$ represents geodesics distance as mentioned in Table \ref{tab1}, and the $w({e_u^{*,m}}, {e_v^{*,m}})$ is defined as:

\begin{equation}
\begin{split}
w({e_u^{*,m}}, {e_v^{*,m}}) = w'({e_u^{*,m}}) + w'({e_v^{*,m}}).
\end{split}
\label{eq:Eq_16}
\end{equation}
Here the importance of a user-item pair is the sum of the influence of $u$ and $v$. Due to the $w'({e_u^{*,m}})$ and $w'({e_v^{*,m}})$ are the entity level subspace attention, the weights can be pre-trained as follows:

\begin{equation}
\begin{split}
w'(e_{u}^{*,m})=\frac{exp(\alpha_{u}^{m})}{\sum_{i=1}^{M}exp(\alpha_{u}^{i})}, 
w'(e_{v}^{*,m})=\frac{exp(\alpha_{v}^{m})}{\sum_{i=1}^{M}exp(\alpha_{v}^{i})}, \\
\end{split}
\label{eq:Eq_17}
\end{equation}

where the $\alpha_{u}$ and $\alpha_{v}$ are matrices of size 1 × M and each entry stands for the importance of the corresponding subspace.

\begin{equation}
\begin{split}
\alpha_{u} = W[e_{u}^{*,1}\ominus_{\kappa}...\ominus_{\kappa}e_{u}^{*,M}],
\alpha_{v} = W[e_{v}^{*,1}\ominus_{\kappa}...\ominus_{\kappa}e_{v}^{*,M}]. \\
\end{split}
\label{eq:Eq_18}
\end{equation}

In summarize, the multiple spaces fusion method has the following advantages: 

\begin{itemize}
\item Different from existing mixed curvature methods \cite{zhang, Kochurov2020, gu2018learning, wang2021mixed} manually combine various manifolds, we adopt the adaptive curvature spaces $\mathbb{U}_{\kappa}^{n}$ instead of constant curvature spaces, e.g., ${\mathbb{H}_{\kappa}^{n}}$, ${\mathbb{E}_{\kappa}^{n}}$ and $  {\mathbb{S}_{\kappa}^{n}}$.

\item To accurately model the complex structure of the graph, we further integrate each subspace in an attention-enhanced manner from a global perspective.
\end{itemize}

\subsection{Geometry-aware Optimization Strategy}
The margin-based ranking loss has shown to be quite beneficial for non-Euclidean recommendation models \cite{sun2021hgcf, yang2022hicf}. This loss seeks to distinguish user-item pairs up to a specified margin into positive and negative samples, once the margin is satisfied the pairs are regarded as well separated. Specifically, for each user $u$ we sample a positive item $i$ and a negative item $j$, and the margin loss is defined as
\begin{equation}
\begin{split}
{\mathcal{L}(u,i,j) = max(\underbrace{dist^{2}(u,i)}_{pull}-\underbrace{dist^{2}(u,j)}_{push}+m, 0)},
\end{split}
\label{eq:Eq_19}
\end{equation}
where the $dist(\cdot)$ denotes the global distance in unified space as mentioned in equation (\ref{eq:Eq_15}),  $m$ is the margin between $(u,i)$ and $(u,j)$.  As a result, positive items are pulled closer to user while negative items are pushed outside the margin. 

However, it is still unclear how to pick a suitable margin. HGCF~\cite{sun2021hgcf} sets the margin to a constant, this action ignores the exponentially expansive geometric properties of hyperbolic spaces. Furthermore, to emphasize the importance of modeling tail items, HICF~\cite{yang2022hicf} sets a greater margin in the vicinity of the hyperbolic origin and a smaller margin elsewhere. Although this method can alleviate the power-law distribution to a certain extent, it will hurt the model performance. E.g., in hyperbolic spaces, the space capacity of the area near the origin is extremely narrow, thus a smaller margin is beneficial to distinguish highly similar entities. While in areas far away from the origin, we prefer a greater margin because the room is spacious enough. On the other hand, in spherical spaces, the origin area is relatively spacious, but the boundary area is narrow. Therefore, we expect assigning larger margins to the area closer to the spherical origin, and smaller margins to the area farther from the spherical origin.

Motivated by the above considerations, we aim to devise a geometric-aware optimization strategy which can benefit from both hyperbolic and spherical spaces. Paper~\cite{pmlr-v80-sala18a} discovered an intriguing phenomenon: when entities $x$ and $y$ progressively moves away from the origin in hyperbolic space, the ratio $\frac{dist(x,y)}{dist(x,\mathbf{o})+dist(y,\mathbf{o})}$ is getting greater. Obviously, the reason is that the numerator (distance) grows significantly faster than the denominator (radius) in hyperbolic spaces. On the contrary, as entities $x$ and $y$ move away from the spherical origin, the ratio $\frac{dist(x,y)}{dist(x,\mathbf{o})+dist(y,\mathbf{o})}$ will gradually become smaller. Surprisingly, the change trend of this ratio is in line with our expectations for the margin. Inspired by these observations, we carefully design the margin $m_g$ as follows: 
\begin{equation}
\begin{split}
{m_g=\sigma(\frac{dist(u,i)}{dist(u,\mathbf{o})+dist(i,\mathbf{o})})+c},
\end{split}
\label{eq:Eq_20}
\end{equation}
where $\sigma$ is the sigmoid function for normalization, and we maintain the constant $c$ for better parameter tuning. Then the $m_g$ is utilized to replace the $m$ in equation (\ref{eq:Eq_19}). As a result, the margin will increase as it slowly departs from the origin in hyperbolic spaces, and in spherical spaces it's the exact opposite. To summarize, either in the hyperbolic or spherical spaces, our model have the ability to perceive the spatial structures and learn the ideal margin.

It is worth mentioning that HICF~\cite{yang2022hicf} aims to improve the attention of tail items in hyperbolic space, thus the margin $m_h$ is set as follows:
\begin{equation}
\begin{split}
{m_h=\sigma{(dist(u,\mathbf{o})+dist(i,\mathbf{o})- (dist(u,i)})+c}.
\end{split}
\label{eq:Eq_21}
\end{equation}
Therefore, it assigns larger margins in narrower areas and smaller margins in spacious areas, exactly the opposite of what we expect. We will compare the performance of various margins in Section \ref{4.3.1}.

\subsection{Time Complexity Analysis}

As we can see, the layer-wise propagation is the core operation in MCKG. Let $N$ denotes the number of user-item interactions, for the $k$-th layer, the matrix multiplication computational complexity is $O(N \cdot |s(v)| \cdot {d_{k}})$, where $s(v)$ represents the item neighbor sampling size, and $d_{k}$ denotes the current transformation size. As a result, the overall time complexity of our proposed model is $O(\sum_{m=1}^{M} \sum_{k=1}^{K} N \cdot |s(v)| \cdot {d_{k}})$, where $M$ is the number of manifolds, and $K$ denotes the number of graph convolution layers. Parameter setting details will be introduced in Section \ref{paramter setting}.

\section{EXPERIMENTS}
In this section, we introduce the experimental settings, and perform extensive comparative experiments with the state-of-the-art models. Then we conduct comprehensive ablation studies to verify the designs in MCKG and reveal the reasons of its effectiveness. Finally, we visualize the learned embedding and analyze the performance of the recommendation model in data-sparse scenarios.
\subsection{Experimental Settings}

\begin{table*}
\renewcommand{\arraystretch}{1}
\caption{\textbf{Three datasets used in this paper.}}
	\centering
	\begin{tabular}{lccc}
		\hline
		Datasets & Movielens-1M & LastFM & Book-Crossing \\
		\hline
		Users/Items & 6,036/2,347 & 1,872/3,846 & 17,860/14,910 \\
		Interactions & 753,772 & 42,346 & 139,746 \\
		\hline
		KG Entities/Relations & 16,954/32 & 9,366/60 & 25,787/18 \\
		KG triples & 20,195 & 15,518 & 60,787 \\
		\hline
	\end{tabular}
	\label{tab2}
\end{table*}

\subsubsection{Datasets}
We perform extensive experiments on the three real-world datasets to assess the performance of the MCKG algorithm. Table \ref{tab1} summarizes the statistics of datasets.

\textbf{Movielens-1M\footnote{https://grouplens.org/datasets/movielens/}:} This movie dataset has been extensively applied to testify recommendation methods. We adopt the version with one million interactions, and each user has more than 20 ratings.

\textbf{LastFM\footnote{https://grouplens.org/datasets/hetrec-2011/}:} The LastFM online music system, which offers the listening count as weight for each user-item interaction, provided the dataset from a set of about 2,000 users.

\textbf{Book-Crossing\footnote{http://www2.informatik.uni-freiburg.de/~cziegler/BX/}:} This dataset provided by the Book-Crossing website, which contains the evaluation of books by more than 10,000 users.

\begin{figure*}
\centering
\includegraphics[scale=.46, trim={40mm 0mm 0mm 0mm}]{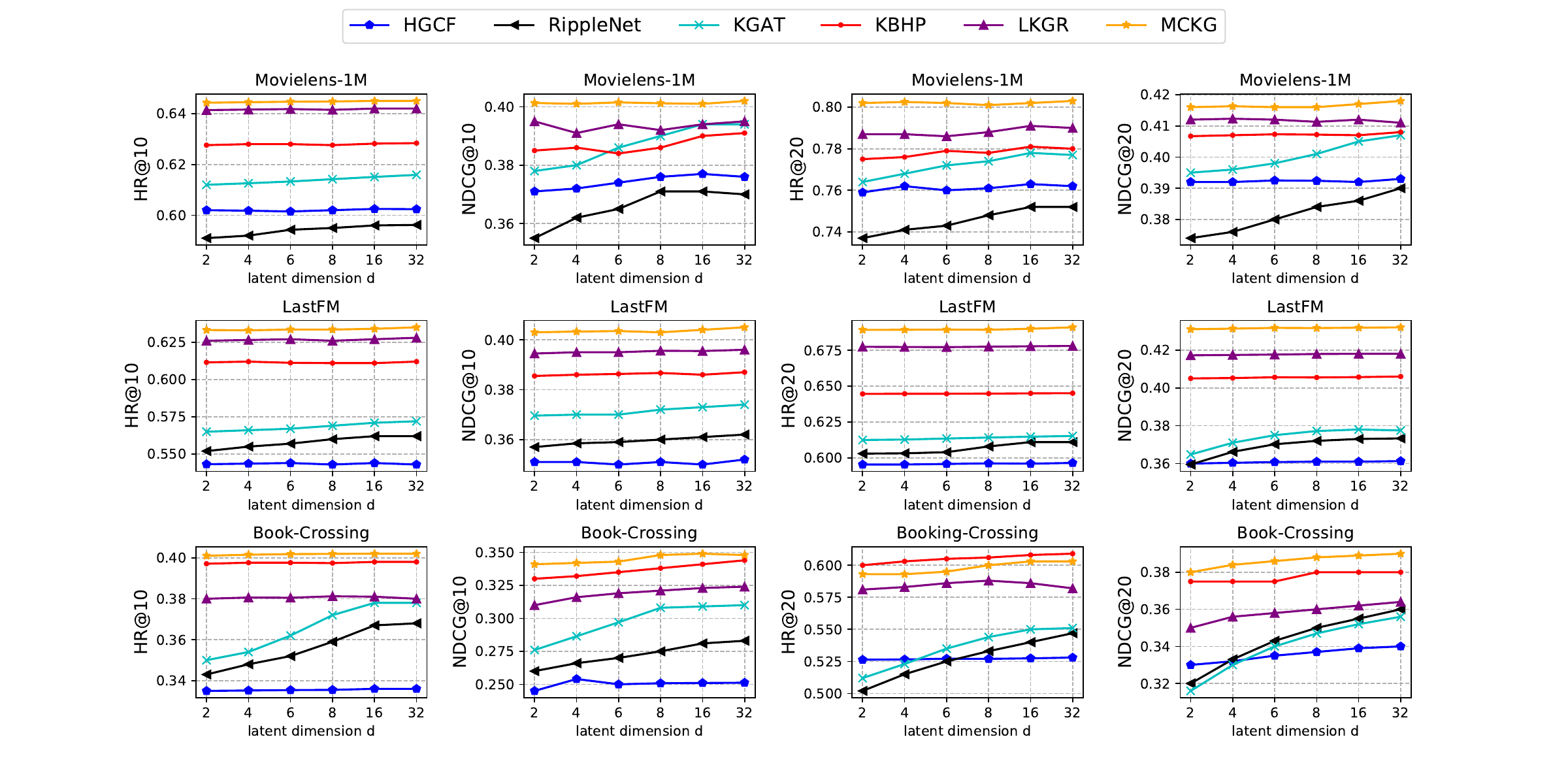}
    \caption{ The performance compared with other competing methods on the three datasets, with latent dimension ranging from 2 to 32.}
    \label{Fig3}
\end{figure*}

\begin{table*}[h]
\renewcommand{\arraystretch}{1.2}
\caption{\textbf{The overall comparison among all competing models, the best results of all methods are in bold, while the second-best results are underlined.}}
\centering
\setlength{\tabcolsep}{1.0mm}{
\begin{tabular}{l|cccc|cccc|cccc}
\hline
{\textbf{Datasets}} &\multicolumn{4}{c|}{\textbf{Movielens-1M}}&\multicolumn{4}{c|}{\textbf{LastFM}}&\multicolumn{4}{c}{\textbf{Book-Crossing}}\\
\cline{2-13}
\textbf{Methods} &\textbf{H@10}& \textbf{H@20} & \textbf{N@10} & \textbf{N@20} & \textbf{H@10} & \textbf{H@20}&\textbf{N@10}& \textbf{N@20} & \textbf{H@10} & \textbf{H@20} & \textbf{N@10} & \textbf{N@20} \\
\hline
\hline
HGCF &0.602 &0.763 &0.376 &0.392 &0.544 &0.596 &0.350 &0.361 &0.336 &0.528 &0.251 &0.340\\
\hline
RippleNet &0.596 &0.752 &0.372 &0.391 &0.562 &0.611 &0.361 &0.372 &0.368 &0.547 &0.281 &0.359\\
\hline
KGAT &0.615 &0.778 &0.394 &0.407 &0.571 &0.614 &0.364 &0.377 &0.379 &0.551 &0.309 &0.356\\

\hline
KBHP &0.628 &0.781 &0.391 &0.408
&0.612 &0.645 &0.387 &0.406
&\underline{0.398} &\textbf{0.609} &\underline{0.344} &\underline{0.390}\\
\hline
LKGR &\underline{0.640} &\underline{0.791} &\underline{0.395} &\underline{0.411}  
&\underline{0.628} &\underline{0.678} &\underline{0.396} &\underline{0.418} &0.380 &0.582 &0.324 &0.364
\\

\hline
MCKG &\textbf{0.645} &\textbf{0.802} &\textbf{0.401} &\textbf{0.418} &\textbf{0.635} &\textbf{0.691} &\textbf{0.405} &\textbf{0.432} &\textbf{0.402} &\underline{0.603} &\textbf{0.348} &\textbf{0.392}  \\
\hline
\hline
Improv. &0.7\% &1.3\% &1.5\% &1.7\% & 1.1\% & 1.9\% & 2.2\% & 3.3\% & 1.0\% &-1.0\% &1.1\% & 0.5\% \\
\hline
\end{tabular}}
\label{tab3}
\end{table*}

\subsubsection{Baselines}
To demonstrate the effectiveness, we compare our proposed MCKG algorithm with the following SOTA methods.
\begin{itemize}

\item HGCF~\cite{sun2021hgcf}: The first work that proposes a hyperbolic GCN model for collaborative filtering. Specifically, HGCF utilizes the Lorentz representation to initialize user and item embeddings, performs graph convolution in the tangent space, and projects back to hyperbolic space to learn the final embeddings.

\item RippleNet~\cite{wang2018ripplenet}: This is the first work to propose the conception of preference propagation. In particular, RippleNet takes the item clicked by the user as the origin, expands around like the water ripples, and constantly absorbs 1-hop and 2-hop neighbors to spread information.

\item KGAT~\cite{wang2019kgat}: 
KGAT trains the embeddings of target node based on its neighbors embeddings, and further recursively executes the embeddings propagation to obtain the high-order connectivity in KG. Additionally, it design an attention mechanism is employed to learn the weight of each neighbor during the information propagation.

\item KBHP~\cite{tai2021knowledge}: KBHP is the first work to combine KG and hyperbolic space for recommendation, which is the hyperbolic version of RippleNet.

\item LKGR~\cite{chen2022modeling}: To better model the scale-free tripartite graphs and investigate the intrinsic hierarchical graph structures, LKGR employs different information propagation strategies in the hyperbolic space to explicitly encode the historical interactions and KG.
\end{itemize}

\subsubsection{Evaluation Metrics} For each user, we randomly pick one item that user has interacted with, and sample 100 unobserved or negative items to construct the test sets. We adopt two common ranking evaluation metrics, i.e., Hit Ratio (HR) and Normalized Discounted Cumulative Gain (NDCG) to evaluate all recommendation methods. As a result, the $HR@K$ clearly assesses whether the test item is appeared on the top-K list, while the $NDCG@K$ assigns higher points to items at the top of the hit list, to emphasize the importance of their positions.

\subsubsection{Parameter Settings}
\label{paramter setting}
In the experiments, we randomly pick $70\%$ of each dataset as the train set, the rest $30\%$ as the test set. For the baseline methods, the model parameters are in accordance with the authors' paper. We set the sampling size and the hop as: 8 and 3 for Movielens-1M and Book-Crossing; 4 and 3 for LastFM. The number of manifolds defaults to 3. Moreover, if HR and NDCG do not increase for 20 successive epochs on the test set, the early stopping strategy is implemented.

\subsection{Overall Performance Comparison}
Table~\ref{tab3} shows the performance (HR@10, HR@20, NDCG@10 and NDCG@20) of all competitive algorithms. The observations obtained from Table~\ref{tab3} are as follows: 
\begin{itemize}
\item Most KG-enhanced methods perform better than HGCF, which proves that the iterative aggregation is an effective way to extract high-order information from KGs. Note that HGCF outperforms RippleNet in the Movielens-1M, one possible reason is that the Movie datasets have a relatively sufficient user-item interactions, so the improvement from KGs is not particularly obvious.
\item Under the same conditions combined with KGs, the hyperbolic model (e.g., KBHP and LKGR) performs better than the Euclidean model (e.g., RippleNet and KGAT). It is crucial to note that KBHP is the hyperbolic version of RippleNet, we can see that in each datasets, KBHP performs markedly better than RippleNet.
\item Compared with other competitive modles, MCKG performs best in most cases. Roughly speaking, this is attributed to our model's ability of adaptive curvature learning and geometry perception. We will reveal this phenomenon in the ablation study.
\end{itemize}

Figure~\ref{Fig3} shows the performance of all the competitive algorithms with different numbers of dimension $d$. With the latent dimension changes, the performance of all models fluctuates within a small range. From this figure, we conclude that Euclidean methods are easily impacted by the embedding size, while hyperbolic methods still perform well even in low dimensions. Specifically, MCKG performs the best in most situations. It is worth mentioning that LKGR also performs relatively well. In addition, KBHP has an excellent performance on Book-Crossing dataset.

\subsection{Ablation Study}
We conduct a comprehensive ablation study on MCKG by showing how the model components affect its performance. 

\subsubsection{Impact of aggregators and geometry-aware margin}
\label{4.3.1}
To further investigate the influence of aggregators (see the equation (\ref{eq:Eq_10}) (\ref{eq:Eq_11}) (\ref{eq:Eq_12})) and margins on the model, we conduct a comprehensive experiment and the results are shown in Table \ref{tab4}, where the suffix '-c' means we adopt constant c as the margin, '-g' indicates the geometry-aware margin in MCKG (see the equation (\ref{eq:Eq_20})), and '-h' represents the margin in HICF (see the equation (\ref{eq:Eq_21})). From this table, we have the following observations:
 
\begin{table}
\renewcommand{\arraystretch}{1.2}
\caption{\textbf{The comparison of performance (HR@20 and NDCG@20) among different variants.}}
\centering
\setlength{\tabcolsep}{1.0mm}{
\begin{tabular}{l|cc|cc|cc}
\hline
{\textbf{Datasets}} &\multicolumn{2}{c|}{\textbf{Movielens-1M}}&\multicolumn{2}{c|}{\textbf{LastFM}}&\multicolumn{2}{c}{\textbf{Book-Crossing}}\\
\cline{2-7}
\textbf{Aggregators} &\textbf{HR}& \textbf{NDCG} & \textbf{HR} & \textbf{NDCG} & \textbf{HR} & \textbf{NDCG} \\
\hline
\hline
GCN-c & 0.783 &0.402 & 0.682 &0.411 &0.534 &0.356\\
\hline
GCN-h & 0.770 &0.393 & 0.675 &0.401 &0.503 &0.332\\
\hline
GCN-g & \textbf{0.802} &\textbf{0.418} & \textbf{0.691} &\textbf{0.432} &0.579 &0.374\\
\hline
\hline
GraphSage-c & 0.754 &0.384 & 0.638 &0.406 &0.576 &0.364\\
\hline
GraphSage-h & 0.730 &0.361 & 0.613 &0.372 &0.568 &0.352\\
\hline
GraphSage-g & 0.771 &0.392 & 0.642 &0.415 &\textbf{0.609} &\textbf{0.392}\\
\hline
\hline
Neighbor-c & 0.742 &0.354 &0.593 &0.351 &0.524 &0.343\\
\hline
Neighbor-h & 0.733 &0.338 &0.561 &0.325 &0.501 &0.327\\
\hline
Neighbor-g & 0.767 &0.365 & 0.602 &0.362 &0.566 &0.366\\
\hline
\end{tabular}}
\label{tab4}
\end{table}

\begin{itemize}

\item The model depends greatly on the aggregator selection. GCN and GraphSage aggregators perform substantially better than Neighbor, one possible reason is that the Neighbor aggregator ignores its own information.

\item All aggregators adopting geometry-aware margins '-g' are significantly better than other margins (i.e., '-c' and '-h'), and the margin performance in HICF is not even as good as a constant margin. This shows that HICF can alleviate the power-law distribution to a certain extent, it will hurt the model
performance.

\item To summarize, the choice of aggregators is still important, and the margin we designed will significantly improve the model performance, especially in larger dataset (e.g., Book-Crossing).

\end{itemize}

\subsubsection{Impact of KG propagation depth}
\begin{table}
\begin{center}
\renewcommand{\arraystretch}{1.2}
\caption{\textbf{The comparison of performance (HR@20 and NDCG@20) among different depths.}}
\centering
\setlength{\tabcolsep}{1.0mm}{
\begin{tabular}{l|cc|cc|cc}
\hline
{\textbf{Datasets}} &\multicolumn{2}{c|}{\textbf{Movielens-1M}}&\multicolumn{2}{c|}{\textbf{LastFM}}&\multicolumn{2}{c}{\textbf{Book-Crossing}}\\
\cline{2-7}
\textbf{depths} &\textbf{HR}& \textbf{NDCG} & \textbf{HR} & \textbf{NDCG} & \textbf{HR} & \textbf{NDCG} \\
\hline
\hline
1 & 0.783 &0.402 &\textbf{0.691} &\textbf{0.432} &0.534 &0.356\\
\hline
2& \textbf{0.802} &\textbf{0.418} &0.684 &0.428 &\textbf{0.609} &\textbf{0.392}\\
\hline
3 &0.787 & 0.407 & 0.662 &0.414 &0.601 &0.376\\
\hline
\end{tabular}}
\label{tab5}
\end{center}
\end{table}

\begin{table}
\begin{center}
\renewcommand{\arraystretch}{1.2}
\caption{\textbf{The comparison of performance (HR@20 and NDCG@20) among different numbers of manifolds.}}
\centering
\setlength{\tabcolsep}{1.0mm}{
\begin{tabular}{l|cc|cc|cc}
\hline
{\textbf{Datasets}} &\multicolumn{2}{c|}{\textbf{Movielens-1M}}&\multicolumn{2}{c|}{\textbf{LastFM}}&\multicolumn{2}{c}{\textbf{Book-Crossing}}\\
\cline{2-7}
\textbf{numbers} &\textbf{HR}& \textbf{NDCG} & \textbf{HR} & \textbf{NDCG} & \textbf{HR} & \textbf{NDCG} \\
\hline
\hline
1 & 0.778 &0.409 &0.663 &0.401 &0.573 &0.356\\
\hline
2 &0.793 & 0.414 & 0.685 &0.430 &0.609 &0.390\\
\hline
3& \textbf{0.802} &\textbf{0.418} &\textbf{0.691} &\textbf{0.432} &0.608 &\textbf{0.392}\\
\hline
4 &0.800 & 0.417 & 0.688 &0.430 &\textbf{0.609} &0.392\\

\hline
\end{tabular}}
\label{tab6}
\end{center}
\end{table}

We consider the aggregation depth from $d = \{1,2,3\}$, the results show that $d=2$ performs best on Movielens-1M, while $d=1$ works best on LastFM and Book-Crossing. We can also see that the model performs best when $d=1$ or $2$, but when $d=3$, all metrics indicate to a rapid collapse. Essentially, the appropriate depth of aggregation  can improve performance, but too long relationship chains are more likely to bring noise.

\subsubsection{Impact of Muti-space Fusion}
We consider the number of manifold from $\{1,2,3,4\}$ and the results are summarized in Table \ref{tab6}, note that there is no space fusion when $number = 1$. Obviously the multi-space fusion is significantly better than the single space. When the $number= \{2,3,4\}$, the performance is stronger than the case of $number = 1$. In addition, the performance of the model is similar in the case of number is 2, 3, or 4. One possible reason is that although the dataset is sparse, the distribution is not particularly complex. As a result, when the number of spaces increases, the performance does not change much.

\begin{figure*}
\centering
  \includegraphics[scale=.5, trim={0mm 0mm 0mm 0mm}]{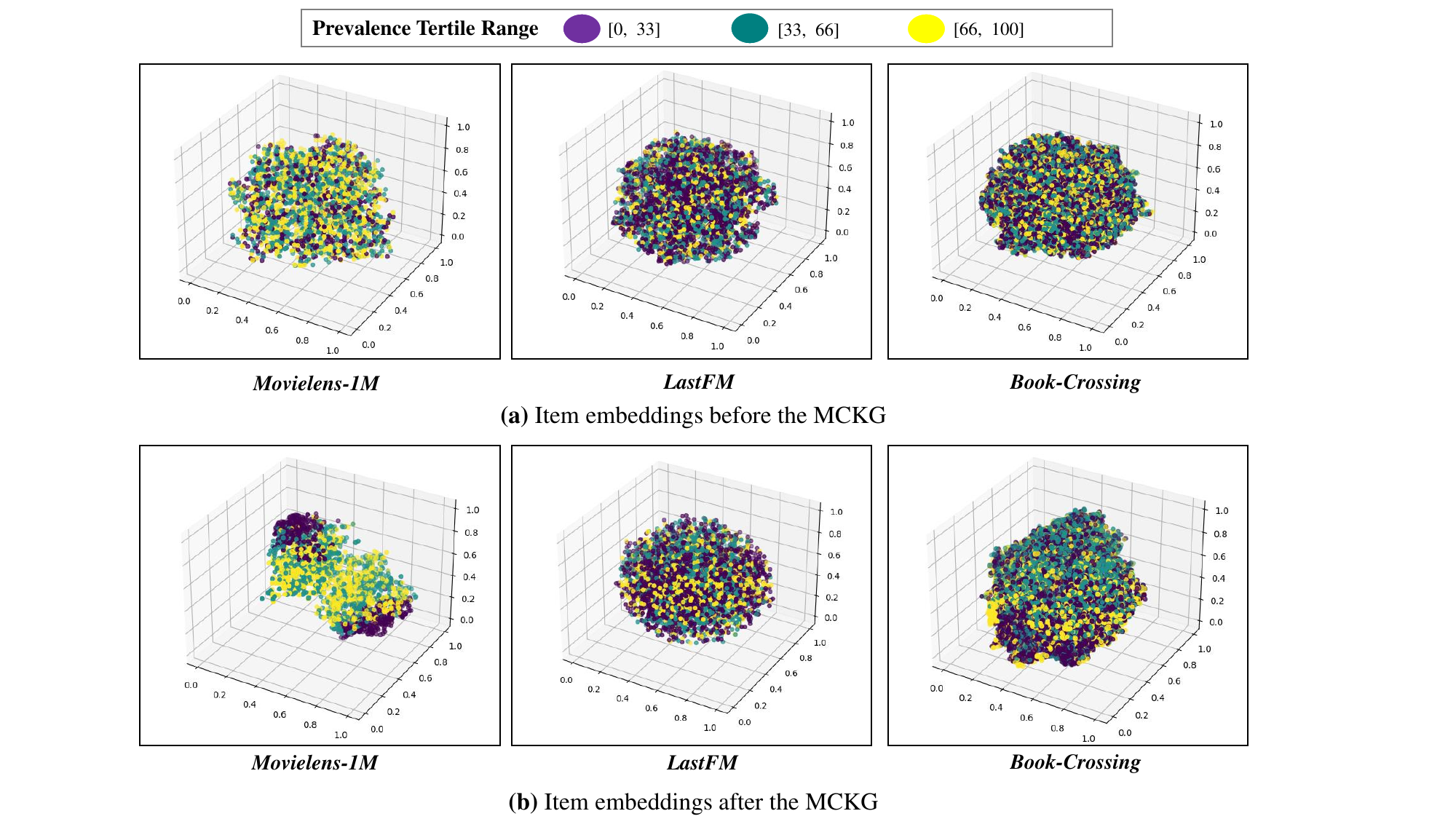}
    \caption{Item embeddings visualisation in the unified space $\mathbb{U}$ before and after the MCKG graph convolutioanl layers.}
    \label{Fig4}
\end{figure*}

\subsection{Embedding Visualization}
We used t-SNE to visualize the item embeddings as shown in Figure \ref{Fig4}. Similar to HGCF~\cite{sun2021hgcf}, to better understand the influence of graph convolutions we depict item embeddings before and after the MCKG layers. Furthermore, the items are divided into three groups based on the prevalence of interactions, and we color each tertile independently. 

Before performing MCKG, we can see that the three groups of item embeddings are evenly distributed in a sphere area. After feeding the three datasets to MCKG, we have the following observations:

\begin{itemize}
\item In Movielens-1M, we can clearly see that unpopular items (purple points) are pushed out at both ends, while the popular items (green and yellow points) are pulled in the middle region. This may be caused by the geometry-aware optimization strategy. Specifically, when the curvature is trainable, our model can distinguish items by popularity, whether in hyperbolic or spherical spaces.

\item In LastFM, we hardly see a clear difference between the trained item embeddings and the original embeddings, but this may be more beneficial for the recommendation's fairness.  On the other hand, in Book-Crossing, we roughly see that the more popular items (green points) are distributed in the upper part of the sphere, and the unpopular items (purple points) are distributed in the lower part of the sphere. 
\end{itemize}

\subsection{Sparse Scene Study}
Data sparsity is a serious problem in  recommendation field, to study the performance of recommendation models in sparse circumstances, we change the training set ratio of Movielens-1M from 10\% to 90\%, the rest datasets as test set. From Table \ref{tab7}, compared with the ratio=90\%, the HR@20 of each model decreases by 6.2\%, 3.7\%, 4.4\%, 2.7\%, 3.0\%, 3.2\% when ratio=10\%. 

With the change of ratio, HGCF fluctuates the most, which shows that compared to other KG-enhanced models, it is more susceptible to sparse data. This also confirms that using KG as auxiliary information is a effective strategy to improve recommendation performance in sparse scenarios. On the other hand, we found that the non-Euclidean methods (i.e., KBHP, LKGR and MCKG) perform more stable than the Euclidean methods (i.e., RippleNet and KGAT) in sparse scene. One possible reason is that due to the spatial characteristics of hyperbolic spaces, even in sparse scene, its ability to distinguish positive and negative items still be maintained.

\begin{table*}
\renewcommand{\arraystretch}{1.2}
\caption{\textbf{Results of HR@20 on MovieLens-1M with different ratios of training set}}
\centering
\begin{tabular}{lcccccccccc}
\hline
\multirow{2}{*}{\textbf{Model}}
&\multicolumn{9}{c}{ \textbf{Ratios of training set }}
&\multirow{2}{*}{\textbf{\%change.}}\\
\cline{2-10}
&\textbf{10\%}&\textbf{20\%}&\textbf{30\%}&\textbf{40\%}&\textbf{50\%}&\textbf{60\%}&\textbf{70\%}&\textbf{80\%}&\textbf{90\%}\\[6pt]

\cline{1-11} 
HGCF & 0.7183 & 0.7320 & 0.7346 & 0.7488 &0.7503 & 0.7538 & 0.7631 &0.7597 &0.7633 &\textbf{6.2\%}\\[6pt] 
RippleNet & 0.7236 & 0.7256 & 0.7263 & 0.7279 & 0.7385 & 0.7492 & 0.7490 & 0.7504 & 0.7403&\textbf{3.7\%}\\[6pt]
KGAT & 0.7456 & 0.7648 & 0.7637 & 0.7646 &0.7648 & 0.7729 & 0.7781 &0.7786 &0.7791 &\textbf{4.4\%}\\[6pt]
KBHP & 0.7611 & 0.7729 & 0.7786 & 0.7865 &0.7878 & 0.7810 & 0.7812 &0.7817 &0.7815 & \textbf{2.7\%}\\[6pt]

LKGR & 0.7690 & 0.7697 & 0.7804 & 0.7806 & 0.7808 &0.7910 & 0.7914 &  0.7905 & 0.7918 &\textbf{3.0\%}\\[6pt]

MCKG & 0.7805 & 0.7809 & 0.7858 & 0.7962 &0.7997 & 0.8023 & 0.8010 &0.8008 &0.8055&\textbf{3.2\%}\\[6pt]
\hline
\end{tabular}
\label{tab7}
\end{table*}

\section{RELATED WORK}
In this section, we review two relevant prior works: KG-enhanced recommendation methods and the hyperbolic representation learning.

\subsection{KG-enhanced Recommendation Methods}
Currently, KG-enhanced recommendation methods has gradually attracted the attention of researchers, here we divide them into embedding-based methods, path-based methods and GCN-based methods: 

Embedding-based approaches typically use knowledge graph embedding (KGE) methods to endow knowledge graph for entity vectors, and further feed these vectors to recommendation module~\cite{cai2018comprehensive}. For instance, to fully exploit the knowledge base, CKE~\cite{zhang2016collaborative} considers three components to extract semantic features from the item's structured content, textual content and visual content, respectively. Finally, CKE is formed to jointly learn the implicit vector of collaborative filtering and the semantics of items based on knowledge base. To dynamically adapt to user preference in real time, DKN~\cite{wang2018dkn} adopts a sequential learning method, that is, the knowledge embedding learning module and the recommendation system module are independent of each other. It uses semantic information and knowledge information to represent news, and makes predictions based on user historical behavior and current candidate news.

Path-based approaches take into account various relationships between items and users in the  knowledge network, which aims to  offer accurate directions for recommendation tasks. Since the KG can construct heterogeneous information networks with user-item interactions, the conventional meta-path techniques can be applied to extracting heterogeneous information networks for recommendation.  For instance, MCRec~\cite{hu2018leveraging} learns representations of users, items and their interaction contexts, i.e., aggregated meta-paths. Specifically, MCRec uses a hierarchical neural network to model the meta-path context as a low-dimensional embedding, and enhances the three representations with a joint attention mechanism. Due to the introduction of meta-path-based context, this model not only has excellent performance, but also has a certain interpretability. 

GCN-based methods typically employ iterative aggregation of the target node's neighbors to fully mine the high-level information of KG. For instance, RippleNet~\cite{wang2018ripplenet} takes the item clicked by the user as the origin, expands around like the water ripples, and constantly absorbs 1-hop and 2-hop neighbors to spread information. To automatically discover the semantic information of KG, KGCN~\cite{wang2019knowledge} samples from the neighborhood of each entity in KG as its receptive field, and then combines the neighborhood information with bias when computing the representation for a given entity. The receptive field can be further extended to multi-hops to model high-order information and capture users' latent long-range interests. Similarly, KGAT~\cite{wang2019kgat} updates a node's embedding based on the embedding of its neighbors, and recursively performs this embedding propagation to capture high-order connectivity in linear time complexity. Additionally, an attention mechanism is employed to learn the weight of each neighbor during propagation.

\subsection{Hyperbolic Representation Learning}
Recently, Hyperbolic representation learning 
has taken an important role in RSs~\cite{vinh2020hyperml, sun2021hgcf, zhang2022geometric, tai2021knowledge, chen2022modeling}. HyperML~\cite{vinh2020hyperml} studies metric learning in hyperbolic space and its connection to CF. Furthermore, HGCF~\cite{sun2021hgcf} proposes a hyperbolic GCN model for CF. To alleviate the power-law distribution in recommender systems, HICF~\cite{yang2022hicf} aims to improve the attention of tail items in hyperbolic spaces, which makes the pull and push process geometric-aware. On the other hand, to reveal the intent factors across geometric spaces, GDCF~\cite{zhang2022geometric} learns geometric disentangled representations associated with user intentions and various geometries. It’s worth emphasizing that KBHP~\cite{tai2021knowledge} and LKGR~\cite{chen2022modeling} are the most similar to our work. Both of them combine KG and hyperbolic representation learning. Different from existing methods that solely adopt a specific manifold, we introduce a unified manifold into KG-based methods to capture the structural information on global level and automatically learn the optimal curvature in the data.

\section{CONCLUSION}

In this work, we proposed a novel recommendation model named MCKG. Our key motivation is that existing non-euclidean recommendation methods have the following issues: 1) existing non-euclidean methods treat $\kappa$ as a trainable parameter, but model parameters constraints depend on the changes of curvature. Once the curvature $\kappa$ varies, the internal structure of manifold may be destroyed, and further unable to obtain high-quality embeddings; 2) Existing mix-curvature methods directly operate on various
manifolds without considering the heterogeneity. To address the above problems, we introduce the unified space can interpolate smoothly
between all geometries of constant curvature, apply multiple spaces fusion to obtain the high-quality embedding, and finally propose a geometry-aware optimization strategy compatible with hyperbolic and spherical spaces. 

In future work, we will further study the mix-curvature method for other application scenarios, e.g., natural language processing ~\cite{liu2021medical, choudhary2022anthem}, computer vision~\cite{ermolov2022hyperbolic, guo2022clipped}, such non-Euclidean modeling strategy learns embeddings of correlated data successfully with reduced distortion.

\begin{acks}
This research work is supported by the National Key Research and Development Program of China under Grant No. 2021ZD0113602, the National Natural Science Foundation of China under Grant Nos. 62176014, 62276015, 62206266, the Fundamental Research Funds for the Central Universities.
\end{acks}

\bibliographystyle{ACM-Reference-Format}
\bibliography{reference}


\begin{thebibliography}{40}


\ifx \showCODEN    \undefined \def \showCODEN     #1{\unskip}     \fi
\ifx \showDOI      \undefined \def \showDOI       #1{#1}\fi
\ifx \showISBNx    \undefined \def \showISBNx     #1{\unskip}     \fi
\ifx \showISBNxiii \undefined \def \showISBNxiii  #1{\unskip}     \fi
\ifx \showISSN     \undefined \def \showISSN      #1{\unskip}     \fi
\ifx \showLCCN     \undefined \def \showLCCN      #1{\unskip}     \fi
\ifx \shownote     \undefined \def \shownote      #1{#1}          \fi
\ifx \showarticletitle \undefined \def \showarticletitle #1{#1}   \fi
\ifx \showURL      \undefined \def \showURL       {\relax}        \fi
\providecommand\bibfield[2]{#2}
\providecommand\bibinfo[2]{#2}
\providecommand\natexlab[1]{#1}
\providecommand\showeprint[2][]{arXiv:#2}

\bibitem[\protect\citeauthoryear{Bachmann, B{\'e}cigneul, and Ganea}{Bachmann
  et~al\mbox{.}}{2020}]%
        {bachmann2020constant}
\bibfield{author}{\bibinfo{person}{Gregor Bachmann}, \bibinfo{person}{Gary
  B{\'e}cigneul}, {and} \bibinfo{person}{Octavian Ganea}.}
  \bibinfo{year}{2020}\natexlab{}.
\newblock \showarticletitle{Constant curvature graph convolutional networks}.
  In \bibinfo{booktitle}{\emph{International Conference on Machine Learning}}.
  PMLR, \bibinfo{pages}{486--496}.
\newblock


\bibitem[\protect\citeauthoryear{Cai, Zheng, and Chang}{Cai
  et~al\mbox{.}}{2018}]%
        {cai2018comprehensive}
\bibfield{author}{\bibinfo{person}{Hongyun Cai}, \bibinfo{person}{Vincent~W
  Zheng}, {and} \bibinfo{person}{Kevin Chen-Chuan Chang}.}
  \bibinfo{year}{2018}\natexlab{}.
\newblock \showarticletitle{A comprehensive survey of graph embedding:
  Problems, techniques, and applications}.
\newblock \bibinfo{journal}{\emph{IEEE Transactions on Knowledge and Data
  Engineering}} \bibinfo{volume}{30}, \bibinfo{number}{9}
  (\bibinfo{year}{2018}), \bibinfo{pages}{1616--1637}.
\newblock


\bibitem[\protect\citeauthoryear{Chami, Ying, R{\'e}, and Leskovec}{Chami
  et~al\mbox{.}}{2019}]%
        {chami2019hyperbolic}
\bibfield{author}{\bibinfo{person}{Ines Chami}, \bibinfo{person}{Zhitao Ying},
  \bibinfo{person}{Christopher R{\'e}}, {and} \bibinfo{person}{Jure Leskovec}.}
  \bibinfo{year}{2019}\natexlab{}.
\newblock \showarticletitle{Hyperbolic graph convolutional neural networks}.
\newblock \bibinfo{journal}{\emph{Advances in neural information processing
  systems}}  \bibinfo{volume}{32} (\bibinfo{year}{2019}).
\newblock


\bibitem[\protect\citeauthoryear{Chen, Wu, Hong, Zhang, and Wang}{Chen
  et~al\mbox{.}}{2020}]%
        {chen2020revisiting}
\bibfield{author}{\bibinfo{person}{Lei Chen}, \bibinfo{person}{Le Wu},
  \bibinfo{person}{Richang Hong}, \bibinfo{person}{Kun Zhang}, {and}
  \bibinfo{person}{Meng Wang}.} \bibinfo{year}{2020}\natexlab{}.
\newblock \showarticletitle{Revisiting graph based collaborative filtering: A
  linear residual graph convolutional network approach}. In
  \bibinfo{booktitle}{\emph{Proceedings of the AAAI conference on artificial
  intelligence}}, Vol.~\bibinfo{volume}{34}. \bibinfo{pages}{27--34}.
\newblock


\bibitem[\protect\citeauthoryear{Chen, Yang, Zhang, Zhao, Meng, Hao, and
  King}{Chen et~al\mbox{.}}{2022}]%
        {chen2022modeling}
\bibfield{author}{\bibinfo{person}{Yankai Chen}, \bibinfo{person}{Menglin
  Yang}, \bibinfo{person}{Yingxue Zhang}, \bibinfo{person}{Mengchen Zhao},
  \bibinfo{person}{Ziqiao Meng}, \bibinfo{person}{Jianye Hao}, {and}
  \bibinfo{person}{Irwin King}.} \bibinfo{year}{2022}\natexlab{}.
\newblock \showarticletitle{Modeling Scale-free Graphs with Hyperbolic Geometry
  for Knowledge-aware Recommendation}. In \bibinfo{booktitle}{\emph{Proceedings
  of the Fifteenth ACM International Conference on Web Search and Data
  Mining}}. \bibinfo{pages}{94--102}.
\newblock


\bibitem[\protect\citeauthoryear{Choudhary, Rao, Katariya, Subbian, and
  Reddy}{Choudhary et~al\mbox{.}}{2022}]%
        {choudhary2022anthem}
\bibfield{author}{\bibinfo{person}{Nurendra Choudhary}, \bibinfo{person}{Nikhil
  Rao}, \bibinfo{person}{Sumeet Katariya}, \bibinfo{person}{Karthik Subbian},
  {and} \bibinfo{person}{Chandan~K Reddy}.} \bibinfo{year}{2022}\natexlab{}.
\newblock \showarticletitle{ANTHEM: Attentive hyperbolic entity model for
  product search}.
\newblock  (\bibinfo{year}{2022}).
\newblock


\bibitem[\protect\citeauthoryear{Ermolov, Mirvakhabova, Khrulkov, Sebe, and
  Oseledets}{Ermolov et~al\mbox{.}}{2022}]%
        {ermolov2022hyperbolic}
\bibfield{author}{\bibinfo{person}{Aleksandr Ermolov}, \bibinfo{person}{Leyla
  Mirvakhabova}, \bibinfo{person}{Valentin Khrulkov}, \bibinfo{person}{Nicu
  Sebe}, {and} \bibinfo{person}{Ivan Oseledets}.}
  \bibinfo{year}{2022}\natexlab{}.
\newblock \showarticletitle{Hyperbolic vision transformers: Combining
  improvements in metric learning}. In \bibinfo{booktitle}{\emph{Proceedings of
  the IEEE/CVF Conference on Computer Vision and Pattern Recognition}}.
  \bibinfo{pages}{7409--7419}.
\newblock


\bibitem[\protect\citeauthoryear{Grattarola, Zambon, Alippi, and
  Livi}{Grattarola et~al\mbox{.}}{2018}]%
        {grattarola2018learning}
\bibfield{author}{\bibinfo{person}{Daniele Grattarola},
  \bibinfo{person}{Daniele Zambon}, \bibinfo{person}{Cesare Alippi}, {and}
  \bibinfo{person}{Lorenzo Livi}.} \bibinfo{year}{2018}\natexlab{}.
\newblock \showarticletitle{Learning graph embeddings on constant-curvature
  manifolds for change detection in graph streams}.
\newblock \bibinfo{journal}{\emph{stat}}  \bibinfo{volume}{1050}
  (\bibinfo{year}{2018}), \bibinfo{pages}{16}.
\newblock


\bibitem[\protect\citeauthoryear{Gu, Sala, Gunel, and R{\'e}}{Gu
  et~al\mbox{.}}{2018}]%
        {gu2018learning}
\bibfield{author}{\bibinfo{person}{Albert Gu}, \bibinfo{person}{Frederic Sala},
  \bibinfo{person}{Beliz Gunel}, {and} \bibinfo{person}{Christopher R{\'e}}.}
  \bibinfo{year}{2018}\natexlab{}.
\newblock \showarticletitle{Learning mixed-curvature representations in product
  spaces}. In \bibinfo{booktitle}{\emph{International Conference on Learning
  Representations}}.
\newblock


\bibitem[\protect\citeauthoryear{Gu, Sala, Gunel, and R{\'e}}{Gu
  et~al\mbox{.}}{2019}]%
        {gu2019learning}
\bibfield{author}{\bibinfo{person}{Albert Gu}, \bibinfo{person}{Frederic Sala},
  \bibinfo{person}{Beliz Gunel}, {and} \bibinfo{person}{Christopher R{\'e}}.}
  \bibinfo{year}{2019}\natexlab{}.
\newblock \showarticletitle{Learning mixed-curvature representations in product
  spaces}. In \bibinfo{booktitle}{\emph{International Conference on Learning
  Representations}}.
\newblock


\bibitem[\protect\citeauthoryear{Guo, Wang, Chen, and Yu}{Guo
  et~al\mbox{.}}{2022}]%
        {guo2022clipped}
\bibfield{author}{\bibinfo{person}{Yunhui Guo}, \bibinfo{person}{Xudong Wang},
  \bibinfo{person}{Yubei Chen}, {and} \bibinfo{person}{Stella~X Yu}.}
  \bibinfo{year}{2022}\natexlab{}.
\newblock \showarticletitle{Clipped Hyperbolic Classifiers Are Super-Hyperbolic
  Classifiers}. In \bibinfo{booktitle}{\emph{Proceedings of the IEEE/CVF
  Conference on Computer Vision and Pattern Recognition}}.
  \bibinfo{pages}{11--20}.
\newblock


\bibitem[\protect\citeauthoryear{Hamilton, Ying, and Leskovec}{Hamilton
  et~al\mbox{.}}{2017}]%
        {hamilton2017inductive}
\bibfield{author}{\bibinfo{person}{William~L Hamilton}, \bibinfo{person}{Rex
  Ying}, {and} \bibinfo{person}{Jure Leskovec}.}
  \bibinfo{year}{2017}\natexlab{}.
\newblock \showarticletitle{Inductive representation learning on large graphs}.
  In \bibinfo{booktitle}{\emph{Proceedings of the 31st International Conference
  on Neural Information Processing Systems}}. \bibinfo{pages}{1025--1035}.
\newblock


\bibitem[\protect\citeauthoryear{Han, Ma, Chen, and Tresp}{Han
  et~al\mbox{.}}{2020}]%
        {han2020dyernie}
\bibfield{author}{\bibinfo{person}{Zhen Han}, \bibinfo{person}{Yunpu Ma},
  \bibinfo{person}{Peng Chen}, {and} \bibinfo{person}{Volker Tresp}.}
  \bibinfo{year}{2020}\natexlab{}.
\newblock \showarticletitle{Dyernie: Dynamic evolution of riemannian manifold
  embeddings for temporal knowledge graph completion}.
\newblock \bibinfo{journal}{\emph{arXiv preprint arXiv:2011.03984}}
  (\bibinfo{year}{2020}).
\newblock


\bibitem[\protect\citeauthoryear{Helgason}{Helgason}{1978}]%
        {helgason1978differential}
\bibfield{author}{\bibinfo{person}{Sigurdur Helgason}.}
  \bibinfo{year}{1978}\natexlab{}.
\newblock \showarticletitle{Differential geometry, Lie groups and symmetric
  spaces}.
\newblock \bibinfo{journal}{\emph{Math. Surveys Monogr}}  \bibinfo{volume}{83}
  (\bibinfo{year}{1978}).
\newblock


\bibitem[\protect\citeauthoryear{Hu, Shi, Zhao, and Yu}{Hu
  et~al\mbox{.}}{2018}]%
        {hu2018leveraging}
\bibfield{author}{\bibinfo{person}{Binbin Hu}, \bibinfo{person}{Chuan Shi},
  \bibinfo{person}{Wayne~Xin Zhao}, {and} \bibinfo{person}{Philip~S Yu}.}
  \bibinfo{year}{2018}\natexlab{}.
\newblock \showarticletitle{Leveraging meta-path based context for top-n
  recommendation with a neural co-attention model}. In
  \bibinfo{booktitle}{\emph{Proceedings of the 24th ACM SIGKDD international
  conference on knowledge discovery \& data mining}}.
  \bibinfo{pages}{1531--1540}.
\newblock


\bibitem[\protect\citeauthoryear{Huang, Zhao, Dou, Wen, and Chang}{Huang
  et~al\mbox{.}}{2018}]%
        {huang2018improving}
\bibfield{author}{\bibinfo{person}{Jin Huang}, \bibinfo{person}{Wayne~Xin
  Zhao}, \bibinfo{person}{Hongjian Dou}, \bibinfo{person}{Ji-Rong Wen}, {and}
  \bibinfo{person}{Edward~Y Chang}.} \bibinfo{year}{2018}\natexlab{}.
\newblock \showarticletitle{Improving sequential recommendation with
  knowledge-enhanced memory networks}. In \bibinfo{booktitle}{\emph{The 41st
  International ACM SIGIR Conference on Research \& Development in Information
  Retrieval}}. \bibinfo{pages}{505--514}.
\newblock


\bibitem[\protect\citeauthoryear{Kochurov, Ivanov, and Burnaev}{Kochurov
  et~al\mbox{.}}{[n.d.]}]%
        {Kochurov2020}
\bibfield{author}{\bibinfo{person}{Max Kochurov}, \bibinfo{person}{Sergey
  Ivanov}, {and} \bibinfo{person}{Eugeny Burnaev}.}
  \bibinfo{year}{[n.d.]}\natexlab{}.
\newblock \showarticletitle{Are Hyperbolic Representations in Graphs Created
  Equal?}
\newblock  (\bibinfo{year}{[n.\,d.]}).
\newblock


\bibitem[\protect\citeauthoryear{Liu, Nickel, and Kiela}{Liu
  et~al\mbox{.}}{2019}]%
        {liu2019hyperbolic}
\bibfield{author}{\bibinfo{person}{Qi Liu}, \bibinfo{person}{Maximilian
  Nickel}, {and} \bibinfo{person}{Douwe Kiela}.}
  \bibinfo{year}{2019}\natexlab{}.
\newblock \showarticletitle{Hyperbolic graph neural networks}.
\newblock \bibinfo{journal}{\emph{Advances in Neural Information Processing
  Systems}}  \bibinfo{volume}{32} (\bibinfo{year}{2019}).
\newblock


\bibitem[\protect\citeauthoryear{Liu, Li, You, Yang, Fan, and Yu}{Liu
  et~al\mbox{.}}{2021}]%
        {liu2021medical}
\bibfield{author}{\bibinfo{person}{Zheng Liu}, \bibinfo{person}{Xiaohan Li},
  \bibinfo{person}{Zeyu You}, \bibinfo{person}{Tao Yang}, \bibinfo{person}{Wei
  Fan}, {and} \bibinfo{person}{Philip Yu}.} \bibinfo{year}{2021}\natexlab{}.
\newblock \showarticletitle{Medical triage chatbot diagnosis improvement via
  multi-relational hyperbolic graph neural network}. In
  \bibinfo{booktitle}{\emph{Proceedings of the 44th International ACM SIGIR
  Conference on Research and Development in Information Retrieval}}.
  \bibinfo{pages}{1965--1969}.
\newblock


\bibitem[\protect\citeauthoryear{Niepert, Ahmed, and Kutzkov}{Niepert
  et~al\mbox{.}}{2016}]%
        {niepert2016learning}
\bibfield{author}{\bibinfo{person}{Mathias Niepert}, \bibinfo{person}{Mohamed
  Ahmed}, {and} \bibinfo{person}{Konstantin Kutzkov}.}
  \bibinfo{year}{2016}\natexlab{}.
\newblock \showarticletitle{Learning convolutional neural networks for graphs}.
  In \bibinfo{booktitle}{\emph{International conference on machine learning}}.
  PMLR, \bibinfo{pages}{2014--2023}.
\newblock


\bibitem[\protect\citeauthoryear{Peng, Varanka, Mostafa, Shi, and Zhao}{Peng
  et~al\mbox{.}}{2021}]%
        {peng2021hyperbolic}
\bibfield{author}{\bibinfo{person}{Wei Peng}, \bibinfo{person}{Tuomas Varanka},
  \bibinfo{person}{Abdelrahman Mostafa}, \bibinfo{person}{Henglin Shi}, {and}
  \bibinfo{person}{Guoying Zhao}.} \bibinfo{year}{2021}\natexlab{}.
\newblock \showarticletitle{Hyperbolic deep neural networks: A survey}.
\newblock \bibinfo{journal}{\emph{IEEE Transactions on Pattern Analysis and
  Machine Intelligence}}  \bibinfo{volume}{14} (\bibinfo{year}{2021}),
  \bibinfo{pages}{1--28}.
\newblock


\bibitem[\protect\citeauthoryear{Qu, Bai, Zhang, Nie, and Tang}{Qu
  et~al\mbox{.}}{2019}]%
        {qu2019end}
\bibfield{author}{\bibinfo{person}{Yanru Qu}, \bibinfo{person}{Ting Bai},
  \bibinfo{person}{Weinan Zhang}, \bibinfo{person}{Jianyun Nie}, {and}
  \bibinfo{person}{Jian Tang}.} \bibinfo{year}{2019}\natexlab{}.
\newblock \showarticletitle{An end-to-end neighborhood-based interaction model
  for knowledge-enhanced recommendation}. In
  \bibinfo{booktitle}{\emph{Proceedings of the 1st International Workshop on
  Deep Learning Practice for High-Dimensional Sparse Data}}.
  \bibinfo{pages}{1--9}.
\newblock


\bibitem[\protect\citeauthoryear{Sala, De~Sa, Gu, and R{\'e}}{Sala
  et~al\mbox{.}}{2018a}]%
        {sala2018representation}
\bibfield{author}{\bibinfo{person}{Frederic Sala}, \bibinfo{person}{Chris
  De~Sa}, \bibinfo{person}{Albert Gu}, {and} \bibinfo{person}{Christopher
  R{\'e}}.} \bibinfo{year}{2018}\natexlab{a}.
\newblock \showarticletitle{Representation tradeoffs for hyperbolic
  embeddings}. In \bibinfo{booktitle}{\emph{International conference on machine
  learning}}. PMLR, \bibinfo{pages}{4460--4469}.
\newblock


\bibitem[\protect\citeauthoryear{Sala, De~Sa, Gu, and Re}{Sala
  et~al\mbox{.}}{2018b}]%
        {pmlr-v80-sala18a}
\bibfield{author}{\bibinfo{person}{Frederic Sala}, \bibinfo{person}{Chris
  De~Sa}, \bibinfo{person}{Albert Gu}, {and} \bibinfo{person}{Christopher Re}.}
  \bibinfo{year}{2018}\natexlab{b}.
\newblock \showarticletitle{Representation Tradeoffs for Hyperbolic
  Embeddings}. In \bibinfo{booktitle}{\emph{Proceedings of the 35th
  International Conference on Machine Learning}}
  \emph{(\bibinfo{series}{Proceedings of Machine Learning Research},
  Vol.~\bibinfo{volume}{80})}, \bibfield{editor}{\bibinfo{person}{Jennifer Dy}
  {and} \bibinfo{person}{Andreas Krause}} (Eds.). \bibinfo{publisher}{PMLR},
  \bibinfo{pages}{4460--4469}.
\newblock


\bibitem[\protect\citeauthoryear{Skopek, Ganea, and B{\'e}cigneul}{Skopek
  et~al\mbox{.}}{2019}]%
        {skopek2019mixed}
\bibfield{author}{\bibinfo{person}{Ondrej Skopek},
  \bibinfo{person}{Octavian-Eugen Ganea}, {and} \bibinfo{person}{Gary
  B{\'e}cigneul}.} \bibinfo{year}{2019}\natexlab{}.
\newblock \showarticletitle{Mixed-curvature Variational Autoencoders}. In
  \bibinfo{booktitle}{\emph{International Conference on Learning
  Representations}}.
\newblock


\bibitem[\protect\citeauthoryear{Sun, Cheng, Zuberi, P{\'e}rez, and
  Volkovs}{Sun et~al\mbox{.}}{2021}]%
        {sun2021hgcf}
\bibfield{author}{\bibinfo{person}{Jianing Sun}, \bibinfo{person}{Zhaoyue
  Cheng}, \bibinfo{person}{Saba Zuberi}, \bibinfo{person}{Felipe P{\'e}rez},
  {and} \bibinfo{person}{Maksims Volkovs}.} \bibinfo{year}{2021}\natexlab{}.
\newblock \showarticletitle{Hgcf: Hyperbolic graph convolution networks for
  collaborative filtering}. In \bibinfo{booktitle}{\emph{Proceedings of the Web
  Conference 2021}}. \bibinfo{pages}{593--601}.
\newblock


\bibitem[\protect\citeauthoryear{Sun, Cao, Zhao, Wan, Zhou, Zhang, Wang, and
  Zheng}{Sun et~al\mbox{.}}{2020}]%
        {sun2020multi}
\bibfield{author}{\bibinfo{person}{Rui Sun}, \bibinfo{person}{Xuezhi Cao},
  \bibinfo{person}{Yan Zhao}, \bibinfo{person}{Junchen Wan},
  \bibinfo{person}{Kun Zhou}, \bibinfo{person}{Fuzheng Zhang},
  \bibinfo{person}{Zhongyuan Wang}, {and} \bibinfo{person}{Kai Zheng}.}
  \bibinfo{year}{2020}\natexlab{}.
\newblock \showarticletitle{Multi-modal knowledge graphs for recommender
  systems}. In \bibinfo{booktitle}{\emph{Proceedings of the 29th ACM
  International Conference on Information \& Knowledge Management}}.
  \bibinfo{pages}{1405--1414}.
\newblock


\bibitem[\protect\citeauthoryear{Tai, Huang, Huang, and Ku}{Tai
  et~al\mbox{.}}{2021}]%
        {tai2021knowledge}
\bibfield{author}{\bibinfo{person}{Chang-You Tai}, \bibinfo{person}{Chien-Kun
  Huang}, \bibinfo{person}{Liang-Ying Huang}, {and} \bibinfo{person}{Lun-Wei
  Ku}.} \bibinfo{year}{2021}\natexlab{}.
\newblock \showarticletitle{Knowledge Based Hyperbolic Propagation}. In
  \bibinfo{booktitle}{\emph{Proceedings of the 44th International ACM SIGIR
  Conference on Research and Development in Information Retrieval}}.
  \bibinfo{pages}{1945--1949}.
\newblock


\bibitem[\protect\citeauthoryear{Vinh~Tran, Tay, Zhang, Cong, and Li}{Vinh~Tran
  et~al\mbox{.}}{2020}]%
        {vinh2020hyperml}
\bibfield{author}{\bibinfo{person}{Lucas Vinh~Tran}, \bibinfo{person}{Yi Tay},
  \bibinfo{person}{Shuai Zhang}, \bibinfo{person}{Gao Cong}, {and}
  \bibinfo{person}{Xiaoli Li}.} \bibinfo{year}{2020}\natexlab{}.
\newblock \showarticletitle{Hyperml: A boosting metric learning approach in
  hyperbolic space for recommender systems}. In
  \bibinfo{booktitle}{\emph{Proceedings of the 13th International Conference on
  Web Search and Data Mining}}. \bibinfo{pages}{609--617}.
\newblock


\bibitem[\protect\citeauthoryear{Wang, Zhang, Wang, Zhao, Li, Xie, and
  Guo}{Wang et~al\mbox{.}}{2018a}]%
        {wang2018ripplenet}
\bibfield{author}{\bibinfo{person}{Hongwei Wang}, \bibinfo{person}{Fuzheng
  Zhang}, \bibinfo{person}{Jialin Wang}, \bibinfo{person}{Miao Zhao},
  \bibinfo{person}{Wenjie Li}, \bibinfo{person}{Xing Xie}, {and}
  \bibinfo{person}{Minyi Guo}.} \bibinfo{year}{2018}\natexlab{a}.
\newblock \showarticletitle{Ripplenet: Propagating user preferences on the
  knowledge graph for recommender systems}. In
  \bibinfo{booktitle}{\emph{Proceedings of the 27th ACM International
  Conference on Information and Knowledge Management}}.
  \bibinfo{pages}{417--426}.
\newblock


\bibitem[\protect\citeauthoryear{Wang, Zhang, Xie, and Guo}{Wang
  et~al\mbox{.}}{2018b}]%
        {wang2018dkn}
\bibfield{author}{\bibinfo{person}{Hongwei Wang}, \bibinfo{person}{Fuzheng
  Zhang}, \bibinfo{person}{Xing Xie}, {and} \bibinfo{person}{Minyi Guo}.}
  \bibinfo{year}{2018}\natexlab{b}.
\newblock \showarticletitle{DKN: Deep knowledge-aware network for news
  recommendation}. In \bibinfo{booktitle}{\emph{Proceedings of the 2018 world
  wide web conference}}. \bibinfo{pages}{1835--1844}.
\newblock


\bibitem[\protect\citeauthoryear{Wang, Zhao, Xie, Li, and Guo}{Wang
  et~al\mbox{.}}{2019c}]%
        {wang2019knowledge}
\bibfield{author}{\bibinfo{person}{Hongwei Wang}, \bibinfo{person}{Miao Zhao},
  \bibinfo{person}{Xing Xie}, \bibinfo{person}{Wenjie Li}, {and}
  \bibinfo{person}{Minyi Guo}.} \bibinfo{year}{2019}\natexlab{c}.
\newblock \showarticletitle{Knowledge graph convolutional networks for
  recommender systems}. In \bibinfo{booktitle}{\emph{The world wide web
  conference}}. \bibinfo{pages}{3307--3313}.
\newblock


\bibitem[\protect\citeauthoryear{Wang, Wei, Nogueira~dos Santos, Wang,
  Nallapati, Arnold, Xiang, Yu, and Cruz}{Wang et~al\mbox{.}}{2021b}]%
        {wang2021mixed}
\bibfield{author}{\bibinfo{person}{Shen Wang}, \bibinfo{person}{Xiaokai Wei},
  \bibinfo{person}{Cicero~Nogueira Nogueira~dos Santos},
  \bibinfo{person}{Zhiguo Wang}, \bibinfo{person}{Ramesh Nallapati},
  \bibinfo{person}{Andrew Arnold}, \bibinfo{person}{Bing Xiang},
  \bibinfo{person}{Philip~S Yu}, {and} \bibinfo{person}{Isabel~F Cruz}.}
  \bibinfo{year}{2021}\natexlab{b}.
\newblock \showarticletitle{Mixed-curvature multi-relational graph neural
  network for knowledge graph completion}. In
  \bibinfo{booktitle}{\emph{Proceedings of the Web Conference 2021}}.
  \bibinfo{pages}{1761--1771}.
\newblock


\bibitem[\protect\citeauthoryear{Wang, He, Cao, Liu, and Chua}{Wang
  et~al\mbox{.}}{2019a}]%
        {wang2019kgat}
\bibfield{author}{\bibinfo{person}{Xiang Wang}, \bibinfo{person}{Xiangnan He},
  \bibinfo{person}{Yixin Cao}, \bibinfo{person}{Meng Liu}, {and}
  \bibinfo{person}{Tat-Seng Chua}.} \bibinfo{year}{2019}\natexlab{a}.
\newblock \showarticletitle{Kgat: Knowledge graph attention network for
  recommendation}. In \bibinfo{booktitle}{\emph{Proceedings of the 25th ACM
  SIGKDD International Conference on Knowledge Discovery \& Data Mining}}.
  \bibinfo{pages}{950--958}.
\newblock


\bibitem[\protect\citeauthoryear{Wang, He, Wang, Feng, and Chua}{Wang
  et~al\mbox{.}}{2019b}]%
        {wang2019neural}
\bibfield{author}{\bibinfo{person}{Xiang Wang}, \bibinfo{person}{Xiangnan He},
  \bibinfo{person}{Meng Wang}, \bibinfo{person}{Fuli Feng}, {and}
  \bibinfo{person}{Tat-Seng Chua}.} \bibinfo{year}{2019}\natexlab{b}.
\newblock \showarticletitle{Neural graph collaborative filtering}. In
  \bibinfo{booktitle}{\emph{Proceedings of the 42nd international ACM SIGIR
  conference on Research and development in Information Retrieval}}.
  \bibinfo{pages}{165--174}.
\newblock


\bibitem[\protect\citeauthoryear{Wang, Huang, Wang, Yuan, Liu, He, and
  Chua}{Wang et~al\mbox{.}}{2021a}]%
        {wang2021learning}
\bibfield{author}{\bibinfo{person}{Xiang Wang}, \bibinfo{person}{Tinglin
  Huang}, \bibinfo{person}{Dingxian Wang}, \bibinfo{person}{Yancheng Yuan},
  \bibinfo{person}{Zhenguang Liu}, \bibinfo{person}{Xiangnan He}, {and}
  \bibinfo{person}{Tat-Seng Chua}.} \bibinfo{year}{2021}\natexlab{a}.
\newblock \showarticletitle{Learning intents behind interactions with knowledge
  graph for recommendation}. In \bibinfo{booktitle}{\emph{Proceedings of the
  Web Conference 2021}}. \bibinfo{pages}{878--887}.
\newblock


\bibitem[\protect\citeauthoryear{Yang, Li, Zhou, Liu, and King}{Yang
  et~al\mbox{.}}{2022}]%
        {yang2022hicf}
\bibfield{author}{\bibinfo{person}{Menglin Yang}, \bibinfo{person}{Zhihao Li},
  \bibinfo{person}{Min Zhou}, \bibinfo{person}{Jiahong Liu}, {and}
  \bibinfo{person}{Irwin King}.} \bibinfo{year}{2022}\natexlab{}.
\newblock \showarticletitle{Hicf: Hyperbolic informative collaborative
  filtering}. In \bibinfo{booktitle}{\emph{Proceedings of the 28th ACM SIGKDD
  Conference on Knowledge Discovery and Data Mining}}.
  \bibinfo{pages}{2212--2221}.
\newblock


\bibitem[\protect\citeauthoryear{Zhang, Yuan, Lian, Xie, and Ma}{Zhang
  et~al\mbox{.}}{2016}]%
        {zhang2016collaborative}
\bibfield{author}{\bibinfo{person}{Fuzheng Zhang},
  \bibinfo{person}{Nicholas~Jing Yuan}, \bibinfo{person}{Defu Lian},
  \bibinfo{person}{Xing Xie}, {and} \bibinfo{person}{Wei-Ying Ma}.}
  \bibinfo{year}{2016}\natexlab{}.
\newblock \showarticletitle{Collaborative knowledge base embedding for
  recommender systems}. In \bibinfo{booktitle}{\emph{Proceedings of the 22nd
  ACM SIGKDD international conference on knowledge discovery and data mining}}.
  \bibinfo{pages}{353--362}.
\newblock


\bibitem[\protect\citeauthoryear{Zhang, Li, Xie, Wang, Shi, Liu, Sun, Zhang,
  Deng, and Zhang}{Zhang et~al\mbox{.}}{2022a}]%
        {zhang2022geometric}
\bibfield{author}{\bibinfo{person}{Yiding Zhang}, \bibinfo{person}{Chaozhuo
  Li}, \bibinfo{person}{Xing Xie}, \bibinfo{person}{Xiao Wang},
  \bibinfo{person}{Chuan Shi}, \bibinfo{person}{Yuming Liu},
  \bibinfo{person}{Hao Sun}, \bibinfo{person}{Liangjie Zhang},
  \bibinfo{person}{Weiwei Deng}, {and} \bibinfo{person}{Qi Zhang}.}
  \bibinfo{year}{2022}\natexlab{a}.
\newblock \showarticletitle{Geometric Disentangled Collaborative Filtering}.
\newblock  (\bibinfo{year}{2022}).
\newblock


\bibitem[\protect\citeauthoryear{Zhang, Li, Xie, Wang, Shi, Liu, Sun, Zhang,
  Deng, and Zhang}{Zhang et~al\mbox{.}}{2022b}]%
        {zhang}
\bibfield{author}{\bibinfo{person}{Yiding Zhang}, \bibinfo{person}{Chaozhuo
  Li}, \bibinfo{person}{Xing Xie}, \bibinfo{person}{Xiao Wang},
  \bibinfo{person}{Chuan Shi}, \bibinfo{person}{Yuming Liu},
  \bibinfo{person}{Hao Sun}, \bibinfo{person}{Liangjie Zhang},
  \bibinfo{person}{Weiwei Deng}, {and} \bibinfo{person}{Qi Zhang}.}
  \bibinfo{year}{2022}\natexlab{b}.
\newblock \showarticletitle{Geometric Disentangled Collaborative Filtering}. In
  \bibinfo{booktitle}{\emph{Proceedings of the 45th International ACM SIGIR
  Conference on Research and Development in Information Retrieval}}.
  \bibinfo{pages}{80--90}.
\newblock


\end{thebibliography}

\appendix

\end{document}